# Multi-Layered Plasmonic Covers for Comb-Like Scattering Response and Optical Tagging

Francesco Monticone, Christos Argyropoulos, and Andrea Alù[*]

Dept. of Electrical & Computer Engineering, The University of Texas at Austin

[*]alu@mail.utexas.edu

*We discuss the potential of multilayered plasmonic particles to tailor the optical scattering response. The interplay of plasmons localized in thin stacked shells realizes peculiar degenerate cloaking and resonant states occurring at arbitrarily close frequencies. These concepts are applied to realize ultrasharp comb-like scattering responses and synthesize staggered, ideally strong super-scattering states closely coupled to invisible states. We demonstrate robustness to material losses and to variations in the background medium, properties that make these structures ideal for optical tagging.*

Increased interest in sensing, optical imaging/tagging and energy harvesting at the nanoscale has recently fostered significant research efforts on enhanced scattering and absorption properties from nanoparticles and, more in general, on the capability of engineering their scattering and absorption spectra at will. Plasmonic nanostructures [1] are particularly well suited for this purpose, because of their unprecedented control and large enhancement of light-matter interaction and the associated strong and localized resonant effects, enabling new and anomalous optical phenomena [2]. In addition to resonant scattering, a notable example of anomalous response from small particles may be obtained with plasmonic *cloaks* [3]. These covers can dramatically reduce the total scattering cross section

(SCS) of moderately sized objects through a scattering cancellation mechanism based on their local negative polarizability. The concept has been also extended to multi-frequency operation by considering multiple plasmonic layers [4]-[5], which provide further degrees of freedom. In this configuration, complex scattering signatures may be achieved because of the complex interaction among multiple plasmon modes. In this context, it was proven [5] that, for any passive scatterer, a resonant scattering peak always exists between two zeros of any scattering order. This fact is an unavoidable constraint of causality and passivity, and it may be considered the scattering equivalent of Foster's reactance theorem in circuit theory [6]. In other words, a zero (cloaked state) in the scattering response is generally followed by a pole (resonant state) before a new zero occurs along the frequency axis.

In the context of cloaking, this inherent property intrinsically limits the overall achievable bandwidth. However, in a more general scenario this same property may be exploited to our advantage, providing unexplored flexibility to engineer the scattering signature of a composite object. The location of alternating zeros and poles in the scattering spectrum obviously depends on the geometry and material properties of the scattering object, and it can be tailored to a very large degree. We have used this idea to realize a degenerate state between cloaking and resonant scattering [7], resulting in an asymmetric signature analogous to a Fano resonance [8]-[10]. In this Letter we apply this concept to further enrich the scattering spectrum of small nanoparticles, producing a complex frequency dispersion that leads to new functionalities. We show that a single, isotropic and center-symmetric scatterer with subwavelength size can in principle realize multiple Fano-like resonances staggered arbitrarily close along the frequency axis, obtaining a peculiar Fano-comb frequency response with combined ideal *superscattering* and *cloaking* features. This ability may open exciting possibilities for several applications, ranging from sensing to spectroscopy [11]-[13] and optical tagging.

Scattering from a concentric multilayered spherical system can be analyzed with Mie theory [14]-[15]. Using the notation of Ref. [3] and assuming an $e^{-i\omega t}$ time convention, the TM scattering coefficient of order $n$ may be written as $c_n^{TM} = -U_n^{TM}/(U_n^{TM} + iV_n^{TM})$ where the quantities $U_n^{TM}$, $V_n^{TM}$ are obtained by solving appropriate $2(N+1) \times 2(N+1)$ determinants [15], where $N$ is the number of concentric layers surrounding a dielectric core (for further details see [16]). TE coefficients can be easily computed using duality. In the long-wavelength regime, where dipolar scattering dominates ($n=1$), the complex wave interaction with multilayered plasmonic scatterers can be understood by analyzing the quasi-static dispersion conditions for cloaking, $U_1^{TM} = 0$, and resonant scattering, $V_1^{TM} = 0$. In order to quantitatively understand the potential offered by a multilayered plasmonic scatterer, consider a composite nanoparticle with geometry depicted in the inset of Fig. 1(a). A dielectric core with radius $a = 150$ nm and permittivity $\varepsilon = 10\varepsilon_0$ is surrounded by $N=4$ plasmonic layers with same thickness and a linearly modulated plasma frequency, i.e., the plasma frequency of two consecutive layers differs by a fixed quantity $\Delta\omega_p$, corresponding to a difference in permittivity $\Delta\varepsilon_c$ at a given frequency. This may be realized, for instance, by gradually varying the doping level in a semiconductor, as discussed below. Under these assumptions, the geometry of the structure is determined once the aspect ratio $\eta_{c1} = a/a_{c1}$, the permittivity $\varepsilon_{c1}$ of the first plasmonic shell and $\Delta\varepsilon_c$ are chosen. The contours in Fig. 1(a) show the dynamic total SCS of the composite nanoparticle as a function of its aspect ratio and permittivity of the first layer, assuming $\Delta\varepsilon_c = -0.15$. Red and blue curves highlight the quasi-static dispersion of resonance and cloaking conditions for dipolar scattering.

It can be generally shown that, for any fixed value of $\eta_{c1}$, the quasi-static dispersion equations $U_1^{TM} = 0$ and $V_1^{TM} = 0$ each admit $N+1$ solutions for $\varepsilon_{c1}$, equal to the number of spherical interfaces in the composite particle. This is related to the fact that each surface plasmon localized at an interface is responsible for a cloaking-resonance, scattering dip-peak, pair. In our scenario, $N+1=5$ cloaking and resonant branches are available, of which only four are visible in Fig. 1(b), since the last pair occurs for larger negative values of $\varepsilon_{c1}$. As discussed in the following, when the layers have similar plasma frequencies as in Fig. 1, the required alternation between cloaking and resonant states implies that several branches concentrate in the region $\varepsilon_{c1} \approx 0$. In this case, cloaking and resonant conditions merge for very thin ($\eta_{c1} \sim 1$) and thick ($\eta_{c1} \sim 0$) shells, and the dual phenomena of cloaking and resonant scattering strongly interfere within the same particle.

To give an example of a practical implementation of the proposed concept, we assume that the plasmonic layers are made of aluminum-doped zinc oxide (*AZO*) semiconductors [16]-[18], whose frequency dispersion is modeled with a lossless Drude model $\varepsilon_{c1} = \varepsilon_0 \left( \varepsilon_\infty - \omega_p^2 / \omega^2 \right)$ with $\varepsilon_\infty = 3.3$. The plasma frequency may be tailored by the doping level and we assume it as $\omega_p = 2213.2$ THz [18] for the first layer, with $\Delta\omega_p = 0.015\,\omega_p$ for the other shells. In order to intercept the region where branches merge in Fig. 1, we choose a high aspect ratio $\eta_{c1} = 0.95$, corresponding to the horizontal dashed arrow in Fig. 1(a). Kramers-Kronig relations require that permittivity decreases with wavelength in low-loss regions [19], as indicated by the arrow, confirming that, as we change the wavelength of operation, zeros and poles necessarily alternate. In the geometry considered here, the sequence of intercepted zeros and poles translates into a peculiar comb-like scattering signature, shown in Fig. 1(b) (red line). Each numeral number in the two panels corresponds to a specific cloaking or resonant state. Three closely spaced ultranarrow resonant peaks and dips appear around the wavelength of 1500 nm, with huge excursions (more than 30 dB) over a narrow bandwidth. These features are produced by a stack of Fano resonances in which cloaking states act as sub-radiant dark modes of the system. In the points where the dispersion curves in Fig. 1(a) get close, the interaction of a dark cloaking state with a bright resonant mode produces a dipole-dipole Fano-like feature [7],[20]-[21], which is much sharper than conventional dipolar plasmonic resonances [peak I in Fig. 1(b)] and provides several advantages compared to the conventional Fano response [7].

Although the scattering is dominated by the dipolar response, the scattering peaks in Fig. 1(b) are more pronounced than conventional dipolar resonances, because of the contribution of a second-order (quadrupolar) plasmonic resonance aligned with the dipolar peaks [this is particularly evident in point I of Fig. 1(b), red line, but it holds true for all resonance peaks]. We prove in the following that this is an important by-product of the proposed configuration, which automatically supports staggered super-scattering resonances [22]. Next, we introduce realistic losses in the Drude model of each layer, with a collision frequency $\gamma = 2 \cdot 10^{-3} \omega_p$ [18] [black line in Fig. 1(b)]. Absorption affects the total scattering excursion in the comb, as expected, but it preserves the comb-like signature. While the quadrupolar contribution is strongly attenuated by losses, the dipolar comb response is robustly preserved and the inverse bandwidth of the staggered Fano-like resonances corresponds to an effective quality factor over ten times larger than an isolated dipolar resonance.

The remarkably different response of the composite nanoparticle at the peaks and dips of the comb is not only evident in its total SCS, but also in the near-field distribution. In Figs. 2(a) and 2(b), we show the electric field in the E plane (time snapshots of the component parallel to the impinging field) for the particle of Fig. 1 at the cloaking dip IV and the resonant peak V, respectively. The near-field undergoes dramatic modifications in this very narrow frequency range. At the resonant peak $\lambda = 1511$ nm, enhanced resonant fields are confined at the interface between two specific plasmonic layers, and the impinging wave is strongly perturbed around the resonant nanoparticle [Fig. 2(b)]. The inset shows a detail of the field amplitude distribution, demonstrating strong field localization in one of the shells. We have verified that each peak in the comb corresponds to the "activation" of a specific interface of the plasmonic cloak. The power flow around the nanoparticle (time-average Poynting vector) is represented as white streamlines and follows complex paths characterized by optical vortices and saddle points [23]-[24]. The nearest cloaking dip occurs at a wavelength just 5 nm longer, and the field distribution is dramatically modified [Fig. 2(a)], as the scattering is almost totally canceled. The extreme proximity in frequency of these very different scattering states ensures a sharp and deep variation between the "on" and "off" states of the frequency comb, which is ideal for applications that require high selectivity between adjacent channels.

For these applications, it is of paramount importance to understand the sensitivity/robustness of this phenomenon to different design parameters. In particular, if tagging applications are envisioned (e.g., multicolor labeling in optical imaging of biological tissues [25]), the sharp resonances should be robust to variations in the background permittivity $\varepsilon_b$. This is possible to assess by analyzing the quasi-static dispersion relations for cloaking $U_n^{TM} = 0$ and resonant scattering $V_n^{TM} = 0$ [16]. For $N$ plasmonic layers, the general solutions do not have a simple analytical form and the best way to visualize them is a graphical representation as in Fig. 1(a). Nevertheless, if we are interested in specific regions of the parameter space, simple analytical formulas may be derived. Letting the aspect ratio $\eta_{c1}$ approach unity (the limit of thin plasmonic shells, as in our example), the dispersion equation for arbitrary $n-$th TM resonance ($V_n^{TM} = 0$) may be written in the simple form:

$$\left(\varepsilon + \frac{n+1}{n}\varepsilon_b\right)\prod_{m=0}^{N}(\varepsilon_{c1} + m\Delta\varepsilon_c) = 0 \tag{1}$$

and for cloaking ($U_n^{TM} = 0$):

$$(\varepsilon - \varepsilon_b)\prod_{m=0}^{N}(\varepsilon_{c1} + m\Delta\varepsilon_c) = 0. \tag{2}$$

Each dispersion equation evidently admits $N+1$ quasi-static solutions, of which one is independent of the shell properties [first factor in Eqs. (1-2)] and corresponds to resonance and cloaking conditions of a homogeneous particle without cover. Conversely, the other terms in (1-2) only depend on the plasmonic shells. Since these are common factors in the two equations, in this limit of ultrathin shells they correspond to $N$ degenerate cloaking / resonant states, which can be seen in the diagram of Fig. 1(a) at $\eta_{c1} \sim 1$. Each of these solutions corresponds to an ultranarrow Fano resonance occurring exactly at the plasma frequency (zero permittivity) of the particular layer that gets activated, as shown in the inset of Fig. 2(b). These degenerate states are completely independent of the core permittivity and the background medium. In addition, this result is surprisingly independent of $n$, implying that, in the limit of thin layers, each resonant peak in the comb supports the superposition of *all scattering orders at resonance at the same frequency*, and each scattering dip represents a true cloaked state, in which *all scattering orders are suppressed at the same frequency*! In other words, the upper and lower limits on scattering excursion in the comb are in principle unlimited in the ideal lossless scenario, as all scattering orders resonate and get suppressed under the same condition. When losses are considered, larger-$n$ harmonics get more affected, but staggered superscattering and invisible states may be still realistically achieved.

If we consider the opposite extreme $\eta_{c1}$ approaching zero (small core, compared to the plasmonic layers), the dispersion relation for resonant scattering may be written as:

$$\left(\varepsilon + \frac{n+1}{n}\varepsilon_{c1}\right) f_n\left(\varepsilon_{c1}, \Delta\varepsilon_c, \varepsilon_b\right) = 0 \qquad (3)$$

and for cloaking:

$$\left(\varepsilon + \frac{n+1}{n}\varepsilon_{c1}\right) g_n\left(\varepsilon_{c1}, \Delta\varepsilon_c, \varepsilon_b\right) = 0, \qquad (4)$$

where $f_n$ and $g_n$ are two analytic functions. The dispersion equations have a common degenerate solution at $\varepsilon_{c1} = -\frac{n}{n+1}\varepsilon$, corresponding to a Fano resonance that depends only on the core and the first plasmonic layer, sustained by the plasmon mode localized at the inner interface. Its sensitivity to the core permittivity combined with nonlinear effects has been exploited in [7] to realize giant all-optical scattering switches. In our scenario the situation is much richer: equations $f_n(\varepsilon_{c1}, \Delta\varepsilon_c, \varepsilon_b) = 0$ and $g_n(\varepsilon_{c1}, \Delta\varepsilon_c, \varepsilon_b) = 0$ admit $N$ additional solutions for $\varepsilon_{c1}$, which are in general nondegenerate cloaking and scattering conditions. Although these solutions do not have a simple general form, after expanding them in Taylor series for small values of $\Delta\varepsilon_c$ we gain additional insights into the comb signature. In this case, we find again that a solution of each equation lies far away, as in Fig. 1(a), and is not of interest for our purposes. The remaining $N-1$ solutions, however, may be written in the form $\varepsilon_{c1} \simeq x_n \Delta\varepsilon_c$, where $x_n$ is a proportionality factor that coincides for cloaking and resonant scattering. In the limit of small $\Delta\varepsilon_c$ they form quasi-degenerate states corresponding to the three "internal" branches of Fig. 1(a). The continuity of these branches and their necessary alternation ensures that a comb feature arises for any aspect ratio in the limit $\Delta\varepsilon_c \to 0$, essentially independent of the background medium. In Figs. 3(a) and 3(b), we numerically demonstrate the inherent robustness of this phenomenon by varying the background and core permittivity, respectively. Consistent with the previous analysis, the comb is completely insensitive to large variations in the background material, whereas the "external" scattering features strongly depend on it. Interestingly, the overall response is also unaffected by the core permittivity.

This peculiar robustness of the scattering signature is further discussed in [16], and it appears perfectly suited for optical tagging. In fact, the spectrum locally assumes a "digitized" regular shape that can be used to robustly encode bits of information in the structure by tuning the plasma frequency of each layer. This concept is sketched in Fig. 4(a): if a light beam is shone on a nanotag composed of our multilayered nanoparticle, the scattered spectrum will be similar to the one in Fig. 1(b). Scattered light can be collected by a photodetector and processed. Now, imagine identifying a fixed narrow "window" of $L = 3$ frequency channels/bits in our photodetector, as indicated in panels (b), (c) and (d) of Fig. 4. The tag identity will be determined by overlapping the comb-like scattering spectrum with the $L$ channels. By tuning the plasma frequencies of the plasmonic layers, it is possible to record the desired code by aligning poles or zeros in the tag windows, hence encoding the identity of up to $2^L$ different tags. This scheme provides a concept example on how the rich scattering spectrum of multilayered plasmonic particles may be used for optical tagging, similar to radio-frequency identification (RFID) tags. The information is effectively encoded in deeply subwavelength nanoobjects and it is possible to read it by means of a scattering measurement. In a practical scenario, we would encode the nanotag at the time of fabrication, and the specific comb signature will hold without being affected by the surrounding environment, allowing efficient encoding and easy detection, particularly interesting to realize nano-biomarkers.

In this Letter, we have shown that a peculiar Fano-comb scattering spectrum can be realized with isotropic multilayered plasmonic nanoparticles. A practical design at infrared frequencies has been proposed based on thin AZO plasmonic layers, whose plasma frequency may be controlled with the doping level. We envision fabrication of these multi-layered shells with a variety of nanofabrication techniques, including nanoskiving [26] a multilayered semiconductor material with a gradient of doping level. The proposed nanoparticles may also be arranged in planar arrays, whose reflection and transmission coefficients would show a similar comb-like signature, realizing thin metasurfaces for optical filtering, sensing and tagging. As an example, we show in [16] the response of a periodic array of Fano-comb particles for different periods of the square lattice. The dipolar nature of these scattering features provides many advantages, including reasonable robustness to realistic material losses. In addition, the comb response automatically realizes staggered super-scattering resonances, which are difficultly realized in conventional plasmonic nanoparticles [22].

The scattering behavior described in this Letter is completely scalable in frequency, provided that plasmonic materials with a controllable plasma frequency are available or realizable in the considered frequency range. Due to the great design flexibility and the possibility to tailor exotic scattering features, we believe that the realization of Fano-comb particles may provide crucial benefits for several applications, such as improving the resolution in comb-spectroscopy techniques and producing efficient optical tagging biomedical devices. Including nonlinearities in the shells or core material may further extend the impact of this concept, providing dynamic tunability and switching effects.

This work was supported in part by the ONR MURI Grant No. N00014-10-1-0942 and by the AFOSR YIP award No. FA9550-11-1-0009.

**References**

[1]  S. Maier, *Plasmonics: Fundamentals and Applications* (Springer, Berlin, 2007).

[2]  J. A. Schuller, E. S. Barnard, W. Cai, Y. C. Jun, J. S. White, and M. L. Brongersma, *Nat. Materials* **9**, 193–204 (2010).

[3]  A. Alù, N. Engheta, *Phys. Rev. E* **72**, 016623 (2005).

[4]  A. Alù, N. Engheta, *Phys. Rev. Lett.* **100**, 113901 (2008).

[5]  A. Alù, N. Engheta, *New J. Phys.* **10**, 115036 (2008).

[6]  R. M. Foster, *Bell Systems Technical Journal* **3**, 259 (1924).

[7]  C. Argyropoulos, P.Y. Chen, F. Monticone, G. D'Aguanno, and A. Alù, *Phys. Rev. Lett.* **108**, 263905 (2012).

[8]  U. Fano, *Phys. Rev.* **124**, 1866-1878 (1961).

[9]  B. Luk'yanchuk, *et al.*, *Nat. Materials* **9**, 707-715 (2010).

[10] A. E. Miroshnichenko, S. Flach, and Y. S. Kivshar, *Rev. Mod. Phys.* **82**, 2257 (2010).

[11]  P. Del'Haye, A. Schliesser, O. Arcizet, T. Wilken, R. Holzwarth, T. J. Kippenberg, *Nature* **450**, 1214 (2007).

[12]  P. Del'Haye, T. Herr, E. Gavartin, M. L. Gorodetsky, R. Holzwarth, T. J. Kippenberg, *Phys. Rev. Lett.* **107**, 063901 (2011).

[13]  A. Schliesser, N. Picque, and T. W. Hansch, *Nat. Photonics* **6**, 440–449 (2012).

[14]  C. F. Bohren and D. R. Huffman, *Absorption and Scattering of Light by Small Particles* (Wiley, New York, 1983).

[15]  J.J. Mikulski and E.L. Murphy, *IEEE Trans. Ant. Propag.* **11**, 169 (1963).

[16]  See Supplemental Material for further details and physical insight on the Mie theory for the scattering of spherical objects, as well as a thorough discussion on the response of a planar array of Fano-comb nanoparticles, the robustness to perturbations of the design parameters and the suitable materials to implement the proposed concepts.

[17]  A. Boltasseva, H. A. Atwater, *Science* **331**, 290 (2011).

[18]  G. V. Naik, J. Kim, and A. Boltasseva, *Opt. Mat. Express* **1**,1090 (2011).

[19]  L. Landau, E.M. Lifschitz, *Electrodynamics of Continuous Media* (Pergamon Press, Oxford, UK, 1984).

[20]  C. Radloff, N. J. Halas, *Nano Letters* **4**, 1323-1327 (2004).

[21]  S. Mukherjee *et al.*, *Nano Letters* **10**, 2694-2701 (2010).

[22]  Z. Ruan, S. Fan, *Phys. Rev. Lett.* **105**, 013901 (2010).

[23]  C. F. Bohren, *Am. J. Phys.* **51**, 323 (1983).

[24]  A. Alù and N. Engheta, *J. Opt. Soc. Am. B* **24**, A89-A97 (2007).

[25]  R. Hu, K.-T. Yong, H. Ding, P. N. Prasad, and S. He, *Journ. of Nanophotonics* **4**, 041545 (2010).

[26]  Q. Xu, R. M. Rioux, M. D. Dickey, and G. M. Whitesides, *Acc. Chem. Res.* 41, 1566 (2008).

**Figures**

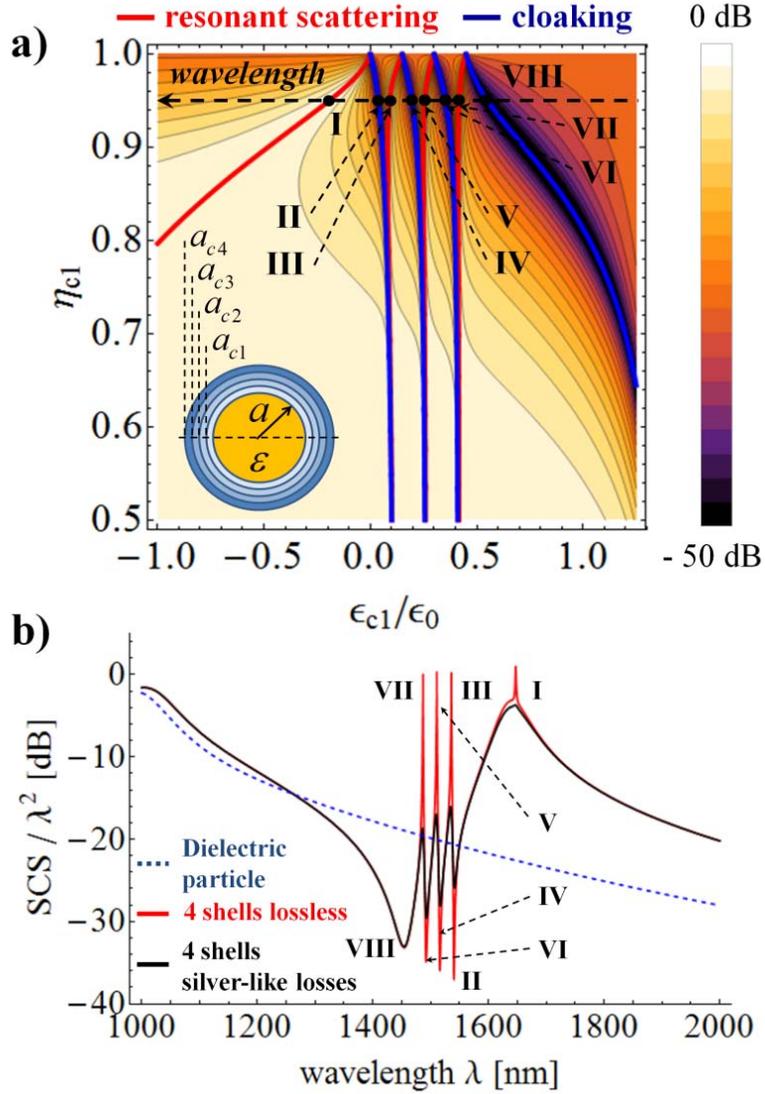

Figure 1 – (a) Total SCS as a function of $\eta_{c1}$ and $\varepsilon_{c1}$ for the multilayered particle shown in the inset, assuming $\varepsilon = 10\varepsilon_0$, $a = 150$ nm and $\Delta\varepsilon_c = -0.15$. The black dashed arrow indicates a wavelength increase for fixed particle geometry, corresponding to panel (b). (b) Scattering spectrum against wavelength for $\eta_{c1} = 0.95$, for lossless (red solid) and lossy layers (black solid) and a homogeneous dielectric particle of same size (blue dashed). Roman numbers identify the different resonant peaks and cloaking dips intercepted when changing the wavelength in (a), reflected in the scattering spectrum in (b).

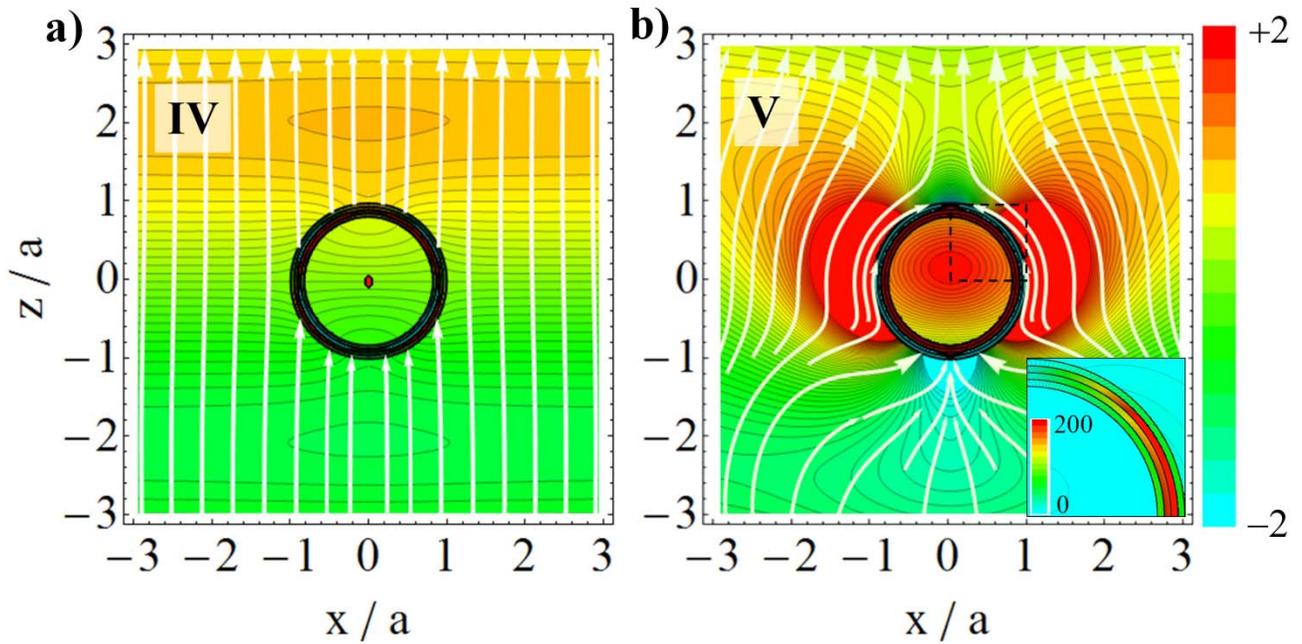

Figure 2 – Electric field distribution in the E plane (snapshot in time), for the multilayered particle of Fig. 1 at the cloaking dip IV (a) and the resonant peak V (b). The power flow (time-average Poynting vector) is shown with white stream lines. The inset in (b) shows the electric field amplitude in the quadrant indicated by the black dashed square.

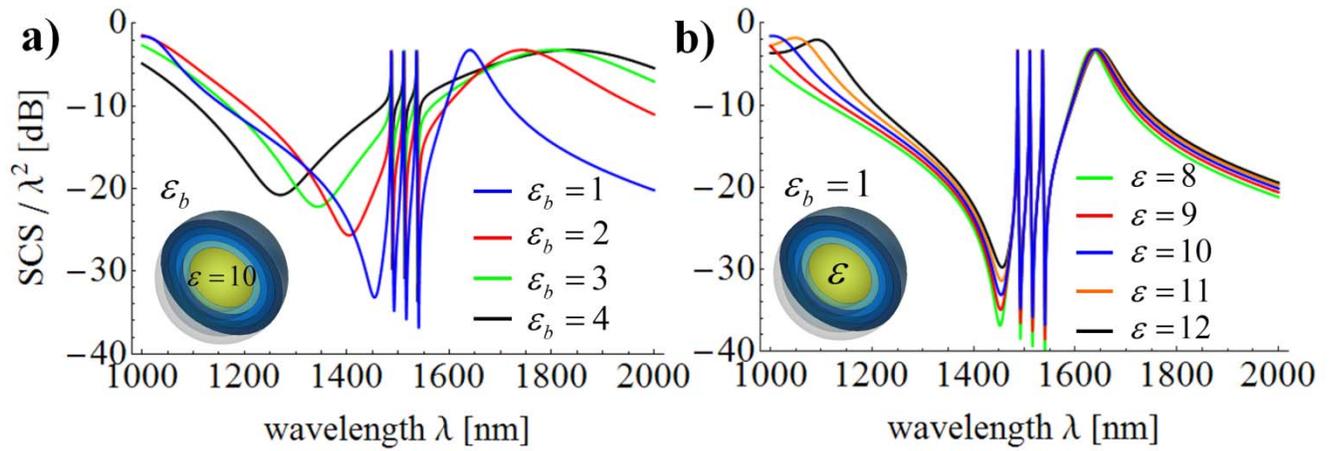

Figure 3 – Sensitivity of the scattering spectrum to the background medium (a) and the core permittivity (b) for a multilayered nanoparticle as in Fig. 1.

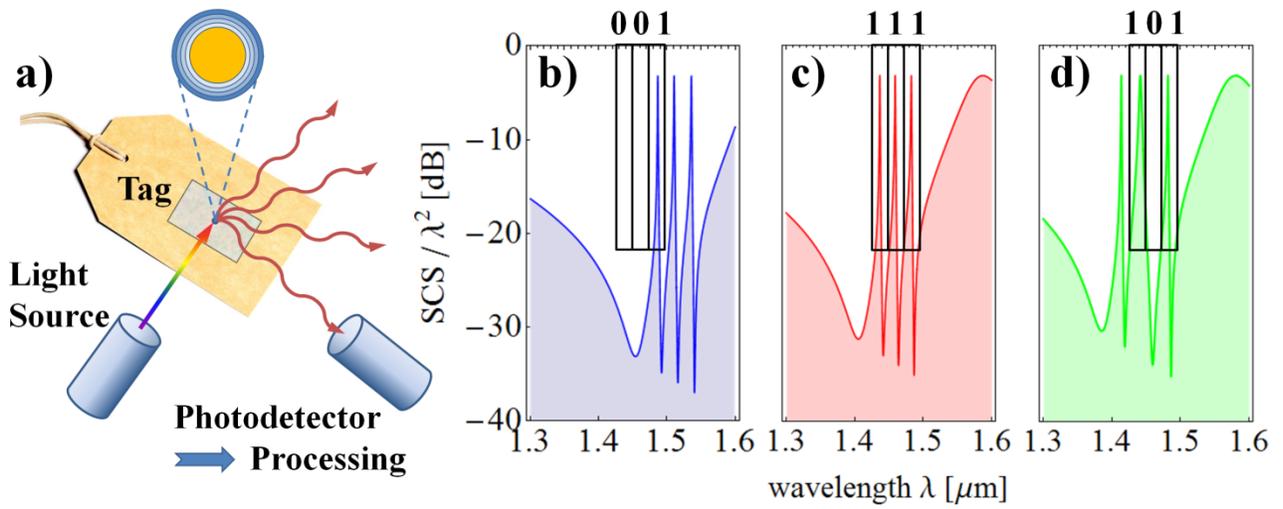

Figure 4 – (a) Schematic representation of optical tag reading. (b),(c),(d) Three examples of optical nanotags: by changing the plasma frequency of the plasmonic layers, it is possible to encode different sequences of bits. The reading window (black boxes) is fixed and by overlapping it with the comb-like scattering response we determine the tag identity.